\documentclass[preprint,eqsecnum,aps,showpacs]{revtex4}
\usepackage{graphicx}
\usepackage{dcolumn}
\usepackage{amsmath}
\input epsf.tex
\begin{document}
\title[]
{Structure peculiarities of cementite and their influence on the 
magnetic characteristics}
\author{A.K. Arzhnikov}
\email{arzhnikov@otf.pti.udm.ru}
\author{L.V. Dobysheva}
\affiliation{
Physical-Technical Institute, Ural Branch of Russian Academy of Sciences, \\
Kirov str.~132, Izhevsk 426001, Russia}

\begin{abstract}
The iron carbide $Fe_3C$ is studied by the first-principle density
functional theory. It is shown that the crystal structure with the carbon
disposition in a prismatic environment has the lowest total energy and
the highest energy of magnetic anisotropy as compared to the structure with
carbon in an octahedron environment. This fact explains the behavior of the
coercive force upon annealing of the plastically deformed samples. The
appearance of carbon atoms in the octahedron environment can be revealed by
Mossbauer experiment.

\end{abstract}
\maketitle
\section{Introduction}

The study of the properties of cementite, that is of iron carbide $Fe_3C$, has
been conducted for a long time. Originally, cementite was investigated
as a component of steels that essentially affects their mechanical
properties. Some attention was paid to the magnetic properties in
connection with nondestructive methods of the steel control. 

A recent splash of interest in the $Fe_3C$ properties may be explained,
first, by new possibilities of obtaining metastable compounds with 
mechanical alloying, implantation and so forth (see, for example, 
Refs.~\cite{Els1,Reed,Koniger,Umemotoa}). Besides, the monophase $Fe_3C$ by
itself has attracted considerable interest due to the peculiarities in
the bulk modulus behavior \cite{Duman1} and the instability of the magnetic
state under pressure \cite{Lin,Duman2}. Finally, an additional impact has
been given by the studies dealing with the chemical composition of the
Earth's core \cite{Wood}.

\begin{figure}[!ht] 
\epsfxsize=10cm
\centerline{\epsfbox{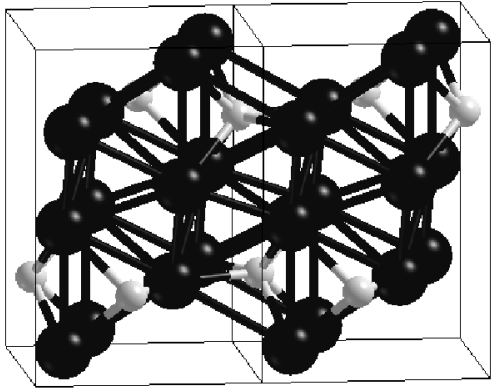}}
\caption{Two unit cells of cementite with prismatic environment
of the carbon atoms. Large black balls show iron atoms, small white are carbon.}
\label{cell-prism}\end{figure}

\begin{figure}[!ht]
\epsfxsize=10cm
\centerline{\epsfbox{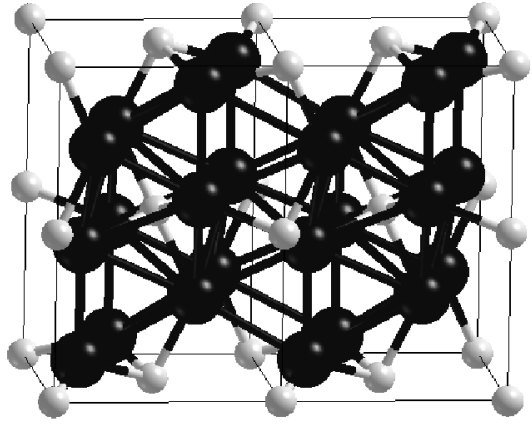}}
\caption{Two unit cells of cementite with octahedron environment
of the carbon atoms.}
\label{cell-octa}\end{figure}

\begin{figure}[!ht] \epsfxsize=9cm
\centerline{\epsfbox{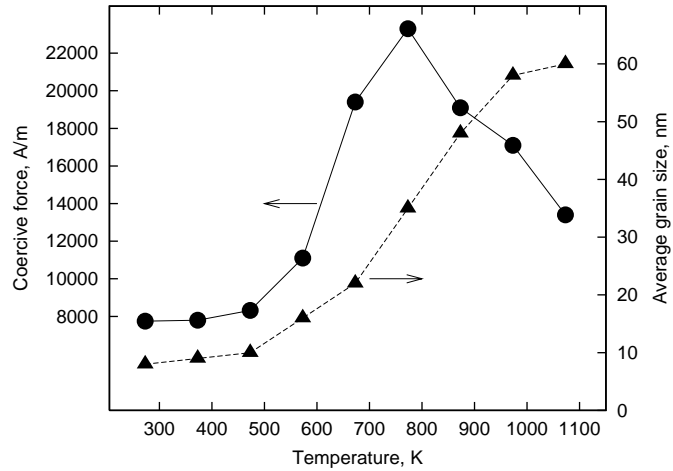}}
\caption{The coercive force and average grain size of cementite as a function 
of annealing temperature \cite{Els2}.}
\hfill{ } \label{coer-f_T}\end{figure}

The cementite structure is believed to be known well enough \cite{Fasiska}
as for the arrangement of iron atoms that is well detected by X-ray
diffraction. The carbon disposition in the lattice is still not clear.
According to Ref.~\cite{Schastl} the carbon atoms can occupy 4 positions
between the iron sites. Two of them (prismatic and octahedron,
Fig.~\ref{cell-prism}  and ~\ref{cell-octa}) were repeatedly specified as
possible places of the carbon atoms. The two other named by authors of
Ref.~\cite{Schastl} distorted prismatic and octahedron positions were not
discussed earlier. Carbon in the two last positions is very close to iron atoms
and these configurations look improbable. The contemporary experiments
cannot still give unequivocal verification of the carbon positions. The
present-day experimantal data on cementite testify only that the carbon
positions actually depend on the mechanical and thermal treatment. The
relevant structural changes manifest themselves in both the mechanical and
magnetic properties. Fig.~\ref{coer-f_T} taken from Ref.~\cite{Els2} shows
the coercive force as a function of the annealing temperature. Its uncommon
behavior may be attributed only to the redistribution of the carbon atoms
in the $Fe_3C$ system: without such a redistribution, annealing should
increase the average grain size (Fig.~\ref{coer-f_T}), reduce the
imperfections of the sample, and decrease thus the coercive force. One can
find other experimental evidence for the changes in the carbon atoms
positions, for example, the change in the number of atoms in the nearest
environment obtained in EELFS \cite{Maratk}, changes in the Mossbauer
spectra \cite{Els3} and so forth.

This paper presents the first-principle calculations conducted in order to
determine the position of carbon and its effect on the physical properties.
Earlier theoretical investigations of cementite have been carried out in
numerous papers (see, for example, \cite{Lin,Medvedeva} and their
references), but only in Ref.~\cite{Medvedeva} a comparison of some
characteristics of the two structures with prismatic and octahedron
positions of carbon was done, the  relaxation of the crystal lattice being
not taken into account.

\section*{Models and methods of calculation}

The crystal structure of $Fe3C$ is an orthorhombic lattice with the lattice
parameters a=0.4523 nm, b=0.5089 nm, c=0.6743 nm \cite{Fasiska}. The unit
cell contains two nonequivalent iron (4 atoms of Fe 1 type and 8 atoms of
Fe 2 type) and one carbon (4 atoms) positions. The closest to iron atoms 
are carbon atoms, in the second sphere of Fe 1 atom there are 5 iron atoms
at slightly different distances, and in that of Fe2 atom there are 8 iron atoms.

The calculations presented in this paper have been conducted  with a
full-potential linearised augmented plane wave (FLAPW) method in the WIEN2k
package \cite{WIEN2k}. In the FLAPW method the wave functions, charge
density and potential are expanded in spherical harmonics within
non-overlapping atomic spheres of radius $R_{MT}$ and in plane waves in the
remaining space of the unit cell (interstitial region).The basis set is
split into the core and valence parts. The core states are treated with the
spherical part of the potential and are assumed to have a spherically
symmetric charge density totally confined inside the muffin-tin spheres;
they are treated in a fully relativistic way. The expansion of the valence
wave functions inside the atomic spheres was confined to $l_{max}=10$ and
they were treated within a potential expanded into spherical harmonics up
to $l=4$. We used an APW+lo-basis \cite{Madsen}. The wave functions in the
interstitial region were expanded in plane waves with a cutoff $K_{max}$
determined by the relation $R_{MT}K_{max}=7$. The charge density was
Fourier expanded up to $G_{max}=20$. First atomic relaxation have been
conducted in the scalar-relativistic approximation for the valence electron
so that the atoms  are in the positions of minimum total energy, where the
calculated forces acting on the nuclei equal zero. A mesh of 60 special
k-points was taken in the irreducible wedge of the Brillouin zone. Using
the unit cell obtained after relaxation, we have conducted calculations of
the magnetic characteristics including the spin-orbit coupling through the
second variational method \cite{WIEN2k} (the number of k-points here was
increased up to 432).

\section*{Results and discussion}

We have calculated periodical lattices with different positions of the
carbon atoms in the unit cell (Fig.~\ref{cell-prism} and 
~\ref{cell-octa}), the lattice parameters being taken equal to the
experimental values. In the unit cell the atoms are shifted to the
positions of minimum energy, that is, the relaxation of the lattice is
made. 

Table~\ref{exp-data} shows that most essential displacements (7 \% for one
distance) occur in the lattice with octahedron environment of the carbon
atoms; the distances between the iron atoms in the lattice with prismatic
position of carbon change by less than 2 \%. 
\begin{table}
\caption{Distances (nm) between the iron atoms (marked as Fe1 and Fe2) 
obtained from X-Ray data (nonrelax) and after the atomic relaxation made 
for two positions of carbon atoms
\label{exp-data}}
\begin{ruledtabular}
\begin{tabular}{c|c|cc}
         &  nonrelax & prismatic & octahedron \\
 Fe2-Fe2 &  0.2495 &  0.2459 &  0.2314  \\
 Fe2-Fe2 &  0.2495 &  0.2521 &  0.2460  \\
 Fe1-Fe2 &  0.2505 &  0.2515 &  0.2521  \\
 Fe2-Fe2 &  0.2543 &  0.2554 &  0.2622  \\
 Fe1-Fe2 &  0.2572 &  0.2597 &  0.2596  \\
 Fe1-Fe1 &  0.2653 &  0.2663 &  0.2629  \\
 Fe2-Fe2 &  0.2653 &  0.2649 &  0.2694  \\
 Fe1-Fe2 &  0.2673 &  0.2649 &  0.2640  \\
 Fe1-Fe2 &  0.2673 &  0.2671 &  0.2686  \\
 Fe1-Fe2 &  0.2700 &  0.2691 &  0.2715  \\
     \end{tabular}
  \end{ruledtabular}
     \end{table}
In spite of the fact that the displacements mightily decrease the energy of
the cell with octahedron position of carbon, the energy of the cell with
prismatic position turns out to be lower, $(E_{oct}-E_{pris}) \approx
0.0075 Ry/at$ (Table~\ref{data-energy}). 
\begin{table}
\caption{Total energy (Ry) of the unit cell for [001] and [010] directions
of magnetisation and its difference $ E_{MA}$ for prismatic and octahedron
environments of carbon atoms
\label{data-energy}}
\begin{ruledtabular}
\begin{tabular}{c|cc}
direction &    prismatic  &  octahedron  \\
    001   & -30852.09422  & -30851.97410 \\
    010   & -30852.09415  & -30851.97411 \\
  \colrule
$E_{MA}$  &       .00007  &      -.00001 \\
     \end{tabular}
  \end{ruledtabular}
     \end{table}
This confirms that the lattice with prismatic carbon position and with
account of the atomic relaxation has a minimum energy, so, it is the ground
state. The difference in energy between the two configurations of the
carbon positions testifies that on the one hand the deformation energy
under mechanical treatment is sufficient to shift the carbon from the
prismatic positions to the octahedron ones, on the other hand the
probability of these shifts through thermodynamic fluctuations is small at
$T < 700 K$. 

The energy with the spin-orbit coupling included depends on the direction
on magnetisation and is given in Table~\ref{data-energy}. 

The magnetic anisotropy energy $E_{MA}$ attracts most attention. Note that for
transition metals it is generally difficult to calculate because of its
small value being a result of the difference of two large quantities  and
close to the calculational inaccuracy. In the case under study, $E_{MA}$ in
the system with prismatic environment of carbon is much higher than in the
system with octahedron carbon position and is calculated more reliably. The
calculations show that magnetizing in the [001] direction is more
preferable than in the [010] one. The magnetic anisotropy energy - the
difference in total energy between the states with magnetization along
these two axes per volume - equals:  
$$ E_{MA} = E_{[010]} - E_{[001]} =
7\times 10^{-5}Ry/cell = 7.9 \times 10^5 J/m^3.  $$ 

The easy-magnetization axis and the magnetic anisotropy energy correspond
to those obtained in the experiment ($E^{ехр}_{MA} = 6.97\times 10^5
J/m^3$) made for a $Fe_3C$ monocrystal at a temperature of 20.4 K
\cite{Blum}. Some distinctions may be referred to the calculational
inaccuracy or to a possible intermixture of carbon in the octahedron
environment in experiment.

For the system with carbon in octahedron environment $E_{MA} < 10^5 J/m^3$
and is seemingly close to the magnetic anisotropy energy of pure iron
$6\times 10^4 J/m^3$. Using a simple model of interacting magnetic moments
at the iron sites along the easy axis, one can find the domain wall width
(see, for example, \cite{Landau}) $\delta=\pi l \sqrt {E_{exch}/E_{MA}}$.
Here $l=0.26 nm$ is the closest distance between the iron atoms, $E_{exch}$
is the exchange energy per volume. Let us estimate $E_{exch}$ from the
temperature of the ferromagnet-paramagnet phase transition ($T_C=480 K$
\cite{Schastl}): $E_{exch} \approx k_B T_C n_{Fe}/(Z V_{cell}) = 5.49\times
10^7 J/m^3$, where $k_B$ is the Boltzmann constant, $n_{Fe}$ is the number
of iron atoms in the unit cell, $Z =11 \div 12$ is the number of nearest
neighbors. So, we obtain $\delta_{pris} \approx 6.1 nm$. Such a domain wall
should be effectively pinned to the defects larger than 10 nm. In
Ref.~\cite{Els2} the samples studied were in the nanocrystalline state with
average grain size of nanocrystals ranging from 10 to 60 nm
(Fig.~\ref{coer-f_T}). It is natural to assume that they are places of the
pinning of the domain walls. In the system with octahedron environment of
carbon, the domain-wall width is $4 \div 5$ times larger ($\delta_{oct} \approx 30
\div 40 nm$), and the nanocrystals lesser than 30 nm are of no significance in
the formation of the coercive force. During annealing of the deformed
cementite, the carbon atoms move from the octahedron positions to the
prismatic ones, the total energy decreasing and the coercive force
increasing (Fig.~\ref{coer-f_T}).

The decrease of the coercive force in the samples after annealing at
temperature higher than 700 K is due to a common mechanism: the degree of
homogeneity of the crystal state becomes higher with annealing temperature
(Fig.~\ref{coer-f_T}).

Note that in spite of the large difference in $ E_{MA}$ between the lattices with
prismatic and octahedron carbon positions the magnitudes of spin or
orbital magnetic moments are not very different in these two lattices (see
Table~\ref{data-magn}). 
\begin{table}
\caption{
Spin $M_{spin}$ and orbital $M_{orb}$ magnetic moments ($\mu_B$),  electric
field gradient $U_{zz}$ ($\times 10^{21} V/m^2$) and asymmetry parameter
$\eta$.
\label{data-magn}}
\begin{ruledtabular}
\begin{tabular}{c|cc|cc}
            &\multicolumn{2}{c|}{prismatic}
	                  &\multicolumn{2}{c}{octahedron}\\
            & Fe1 &  Fe2  &   Fe1  &  Fe2   \\
  \colrule
$M_{spin}$  & 2.01&  1.94 &   1.88 &  1.69  \\
$M_{orb}$   & 0.05&  0.04 &   0.05 &  0.03  \\
  \colrule
$U_{zz}$    & 3.12&  1.34 &   2.94 & -2.37  \\
$\eta$      & 0.12&  0.64 &   0.88 &  0.22  \\
     \end{tabular}
  \end{ruledtabular}
     \end{table}
Taking into account the fact that the samples always contain some amount of
other phases of iron and carbon, the magnetisation measurements do not give
a possibility to distinguish the carbon positions. The difference in $ E_{MA}$
for systems with equal spin moments and equal orbital moments results from
a difference in solely the electron-density distribution.
Table~\ref{data-magn} shows that the electric field gradients $U_{zz}$ at
iron sites in the second inequivalent position are of opposite sign in the
systems with prismatic and octahedron environment of carbon. This leads to
a quadrupole interaction of different sign, and there should exist a
difference in the Mossbauer spectra. The large quadrupole splitting $\Delta
= 0.5 e^2 Q U_{zz} = 1.44\times 10^{-27}J$, ($eQ = 0.18\times 10^{-28}
m^2$, see Ref.~\cite{Kienle}) and the large share of these atoms in the
cell $n_{Fe2} = 8$ allow one to believe that a difference in the shape of the
Mossbauer spectrum may be experimentally observed for the carbon atoms in an
octahedron or a prismatic position. Difficulties in interpreting such
Mossbauer experiments arise from the simultaneity of the electric and
magnetic interactions. The combined hyperfine interaction, the angle
between the hyperfine magnetic field and the electric field gradient, and
the spatial averaging may essentially complicate the Mossbauer spectrum
shape. 

\section*{Conclusions}

With the help of the first-principle calculations we show that the magnetic 
anisotropy energy of cementite is much higher when the carbon atoms are in 
the prismatic pores in contrast with the structure when the carbon atoms 
occupy the octahedron pores. The former structure has a lowest total
energy. These calculations explain the experimental behavior of coercive
force as a function of the annealing temperature for the plastically
deformed samples. The experimental data and the results of calculations 
confirm a
possibility of different carbon disposition between the iron sites in
cementite and the movement of carbon atoms during the mechanical or
thermal treatment. Such structural changes can be directly detected in the
Mossbauer experiment by a change in quadrupole splitting.

\section*{Acknowledgments}

The authors are grateful to Prof. E.~P.~Yelsukov and Prof. A.~I.~Ul'yanov 
for helpful discussions and for experimental data. This work was partially
supported by INTAS (grant 03-51-4778), and RFBR (grant 06-02-16179).

\end{document}